\documentclass[11pt]{article}

\usepackage{newtxtext,newtxmath}

\usepackage{graphicx}

\usepackage[letterpaper,margin=1in]{geometry}

\usepackage{float}
\usepackage{placeins}
\usepackage{setspace}
\usepackage{xcolor}

\definecolor{PNASorange}{RGB}{200,89,45}

\renewenvironment{abstract}
	{\quotation}
	{\endquotation}

\date{}

\makeatletter
\renewcommand{\fnum@figure}{\textbf{Figure \thefigure}}
\renewcommand{\fnum@table}{\textbf{Table \thetable}}
\makeatother

\usepackage{scicite}

\usepackage{url}

\newcommand{\mytexttilde}{\raisebox{0.5ex}{\texttildelow}}

\def\scititle{
Noninvasive precision modulation of high-level neural population activity via natural vision perturbations}
\title{
  {\Huge \bfseries \boldmath %
  \begin{spacing}{1}
    \scititle
  \end{spacing}}
}

\author{
	Guy~Gaziv$^{12\dagger}$ \hspace{.5cm}
	Sarah~Goulding$^{1}$ \hspace{.5cm}
        Ani~Ayvazian-Hancock$^{1}$ \hspace{.5cm}
        Yoon~Bai$^{1}$ \hspace{.5cm}
        James~J.~DiCarlo$^{123}$ \vspace*{.2cm} \and
	\small$^{1}$McGovern Institute for Brain Research, Dept. of Brain and Cognitive Sciences, Massachusetts Institute of Technology\and
        \small$^{2}$Center for Brains, Minds, and Machines, MIT\and
        \small$^{3}$MIT Quest for Intelligence\and
	\small$^{\dagger}$Correspondence: guyga@mit.edu
}

\begin{document} 

\maketitle

\vspace*{-.5cm}

\begin{abstract} \bfseries \boldmath

Precise control of neural activity -- modulating target neurons deep in the brain while leaving nearby neurons unaffected -- is an outstanding challenge in neuroscience, generally approached using invasive techniques. This study investigates the possibility of precisely and noninvasively modulating neural activity in the high-level primate ventral visual stream via perturbations on 
one's natural visual feed. When tested on macaque inferior temporal~(IT) neural populations, we found quantitative agreement between the model-predicted and biologically realized effect: strong modulation concentrated on targeted neural sites. We extended this to demonstrate accurate injection of experimenter-chosen neural population patterns via subtle perturbations applied on the background of typical natural visual feeds. These results highlight that current machine-executable models of the ventral stream can now design noninvasive, visually-delivered, possibly imperceptible neural interventions at the resolution of individual neurons.

\end{abstract}

\vspace*{.5cm}

\begin{figure}[t]  %
  \centering
  \vspace*{-.5cm}
  \includegraphics{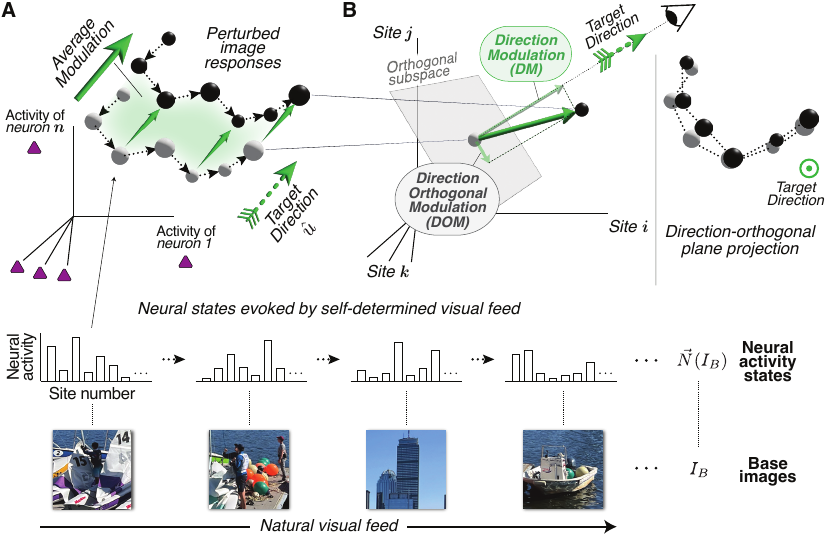}
  \caption{\textbf{Framework for defining and evaluating directional neural population modulation.}
  {\small {\it
  (\textbf{A})~During natural vision, a series of self-determined fixations produces a series of images on the central retinea (here called ``base images''; lower panel) that drive the IT neural population into a series of activity states~(gray dots in~(A)).  Given an arbitrary experimenter-chosen, predetermined direction defined in IT neural population space, the ``non-interrupting'' modulation goal is to consistently displace the IT population such that the IT neural activity state is shifted strongly along the intended modulation direction~(a.k.a. target direction; green arrow), with ideally no shift in any other direction~(see~(B)).  And this ideally needs to hold true for any potential input (base) image, so that it might be continuously applied during natural vision.  
  (\textbf{B})~Precision is evaluated for each base image via on/off axis decomposition of the effect of an applied image perturbation. Efficient image perturbations are those that drive strong Direction Modulation~(DM) with little to no Direction Orthogonal Modulation~(DOM). These constraints are directly imposed as objectives to guide the model-based image perturbations.  And the same measures~(DM and DOM) are used to evaluate the empirical effects in the monkey physiology experiments.
  }}
  }
  \vspace*{-.2cm}
  \label{fig:Goal}
\end{figure}

\begin{figure}[t]  %
  \centering
  \vspace*{-.4cm}
  \includegraphics{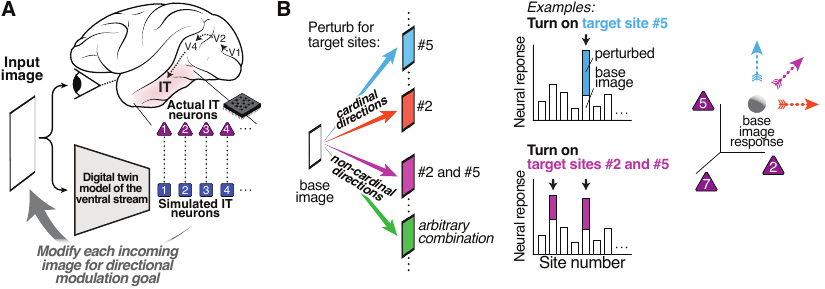}
  \caption{\textbf{Model-guided experiments to test directional neural population modulation.}
  {\small {\it
  (\textbf{A})~On Day 1, we record macaque IT cortex neural multi-unit responses to a set of natural images. Overnight, we use those responses to map weighted combinations of ventral-stream model IT neurons to a sub-population of visually-reliable biological IT neural sites~(typically~10 sites). The simulated neural population is then used in a computer to guide the construction of perturbations to a set of base images~(not used for mapping) for one or more directional modulation goals~(see panel (B)). The neural responses to the base and the perturbed images are recorded on the next day. We compare the model-predicted and monkey-measured modulation results. 
  (\textbf{B})~Examples of possible modulation goals. Each base image can, in theory, be perturbed in service of infinitely many potential IT-directional modulation goals. For simplicity, we focus on two families of modulation goals: cardinal directional modulation, where only a single IT site is targeted with the goal of keeping the other recorded IT sites unaffected, and non-cardinal directions, involving all otherwise convex combinations of IT sites.
  }}
  }
  \vspace*{-.2cm}
  \label{fig:Approach}
\end{figure}

\begin{figure}[!ht]
  \centering
  \includegraphics{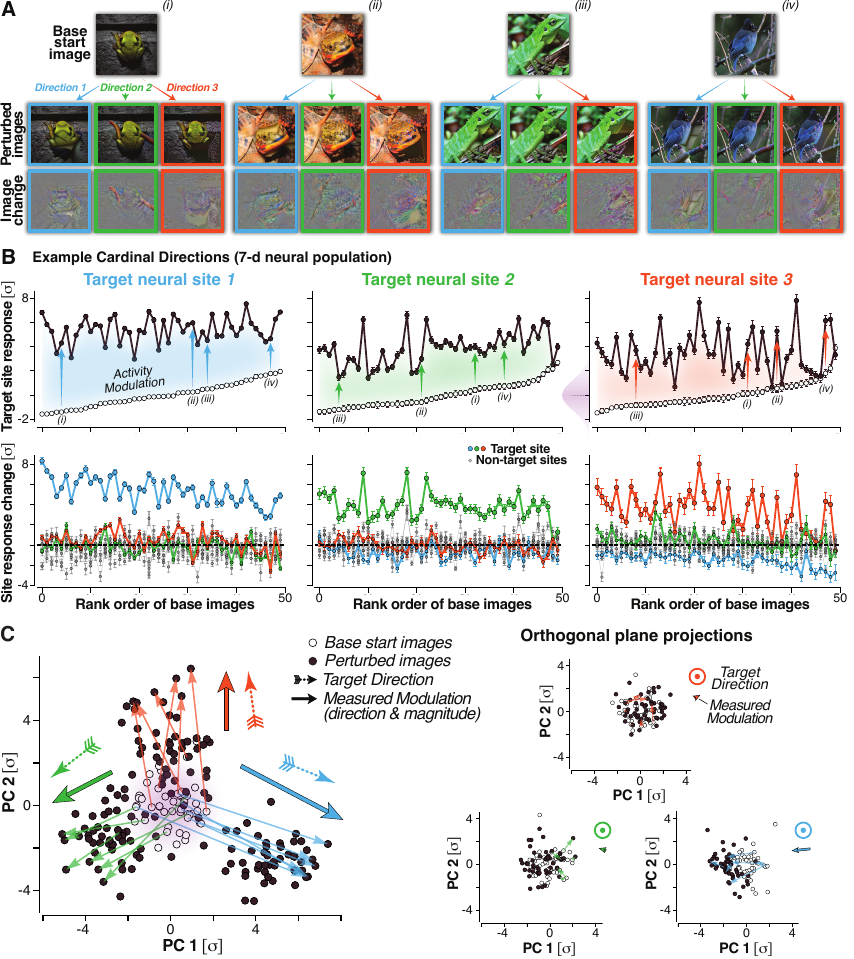}
  \vspace*{-.2cm}
  \caption{\textbf{Cardinal modulation examples.}
    {\small {\it
    We designed perturbed images to try to increase the activity of a target neural site while not affecting the others.
    Showing results for three target sites in a 7-site neural population for~50 base images.
  (\textbf{A})~Examples of base images and their perturbed versions for three cardinal directions selected. 
  (\textbf{B})~Top: Neural responses to all base images~(rank ordered; open dots) and their perturbed versions~(filled dots) for each targeted site. 
  Average modulation size is as large or larger than the site's natural operating range~(i.e., range of the open dots).
  Roman letters indicate example images. 
  Bottom: Similar but showing activity \textit{change} relative to the base image responses for target and non-targeted sites~(color coded).
  Error bars, SEM by bootstrap over 45 image presentation repeats.
  (\textbf{C})~PCA visualizations of the neural responses in the plane of the three intended directions~(left, see Main Text) and in the orthogonal planes~(right). Average modulation arrows are to scale. 
  }}
  }
  \vspace*{-.2cm}
  \label{fig:ConceptualResults}
\end{figure}

\noindent

A goal of applied systems neuroscience is to develop technologies that precisely intervene on ongoing neural activity for both scientific and applied goals. One such ideal neural modulation technology would provide the capability of selectively modifying the activity of \emph{any} predetermined set of target neurons in a given brain area, each by a prescribed amount -- without interrupting the normal role of all \emph{other} nearby neurons in that area.

For brain areas supporting visual processing, such desired ``non-interrupting'' technology would, for example, elevate the activity of targeted neurons, while preserving the response patterns evoked by the natural ongoing visual input of non-targeted neurons. Geometrically, this corresponds to precisely and continuously displacing the image-evoked neural population state along a predetermined target \emph{direction} in the neural population space~(a unit vector $\hat{u}$; Fig.~\ref{fig:Goal}A). The technology would ideally support prescribed displacements of potentially high amplitude, and with direction perfectly aligned with the target direction, \emph{regardless} of visual input.

Possible non-interrupting technologies include invasive intra-cranial techniques, where neural activity is directly elevated via electrical or light-induced interfaces~\cite{Boyden2005Millisecond-timescaleActivity, Ditterich2003MicrostimulationDecisions, Salzman1990CorticalDirection, Yizhar2011NeocorticalDysfunction,Chen2018Near-infraredOptogenetics}. But their precision remains crude~\cite{Jazayeri2017}. Another option is to find molecules on the target neurons and then develop pharmaceutical/molecular targeting of those. However, targets may not exist for arbitrary target neurons, and delivery and safety challenges loom large.

In this study, we explore an alternative, noninvasive approach that seeks to ride alongside the natural visual processing pathways to achieve precision neural modulation~(Fig.~\ref{fig:Goal}). Specifically, we test the possibility of using perceptually subtle perturbations of the naturally occurring spatial energy pattern impinging on the sensory epithelium, here the retinae. 

A key observation supporting this approach is that, over the past decade, certain artificial neural network~(ANN) models of the ventral stream~\cite{Kar2024,Yamins2016UsingCortex,Kriegeskorte2015DeepProcessing} are accurate enough to guide the generation of de novo synthetic stimuli to ultra-activate targeted neural sites in an intermediate region of the primate ventral stream~\cite{Bashivan2019NeuralSynthesis}. And some of these sensory computable models have been successfully tasked to discover small magnitude pixel changes that cause 
large neural changes in targeted high level neurons~\cite{Guo2022AdversariallyRepresentations} -- demonstrating that they can ``see'' epithelial modulation avenues that were previously invisible to us.

Are these models accurate enough to be used as engine technology for epithelially-delivered, precision modulation of neural population activity in a deep visual area? 
Specifically, using neurophysiology experiments in visually fixating Rhesus macaques, we asked, can we use current models to design image perturbations that induce strong precision neural modulation for arbitrary experimenter-chosen goals, defined over recorded sub-populations at the highest stage of ventral visual processing, the inferior temporal~(IT) cortex~\cite{DiCarlo2012HowRecognition}?

We defined the achieved neural population \textit{modulation} as the change vector of the perturbed-image-evoked IT neural population activity state relative to the ``background'' IT population state evoked by the clean~(unperturbed) image~(Fig.~\ref{fig:Goal}A). In typical natural viewing settings, wherein natural scenes are sampled through saccadic eye movements~(Fig.~\ref{fig:Demo}A), the background state updates with each new visual input, yielding a set of change vectors corresponding to each base image. To evaluate the goodness of an intended modulation, we decompose each measured change vector into two components~(Fig.~\ref{fig:Goal}B): (i)~its amplitude \emph{along} the target direction vector, referred to as \textbf{Direction Modulation}~(DM, a scalar), and (ii)~its projection on the hyperplane orthogonal to the target vector, referred to as \textbf{Direction Orthogonal Modulation}~(DOM, a vector).
An ideal non-interrupting modulation technology should, for any predetermined target direction, enable achieving a large DM~(``intended effect'') with zero DOM magnitude~(``side effects'').

Fig.~\ref{fig:Approach} overviews our model-based approach implementing the proposed neural modulation technology. 
We used a two-day experimental, closed-loop protocol. On protocol Day 1, we recorded the activity rates of a population of~(typically ten) biological IT neural sites in response to a set of natural images. 
Overnight, we used those response data to map~(i.e., regress) artificial neurons in the IT layer of our leading ANN model of the ventral stream to that set of biological-IT neural sites~\cite{methods}.  
We then used the model to guide the design of a set of image perturbations -- pixel-level changes, per image -- for a new set of base~(clean) natural images, which were not used during mapping.  
On Day 2, we recorded neural responses to both the base and perturbed images for the same set of IT sites, allowing us to compute a perturbation-induced change vector~(as above) for each base image. This two-day protocol, which defines a single experiment, was used to test a range of directional modulation goals, and the results presented include six experiments\footnote{In Main Text alone.} in three monkeys.

A plethora of predetermined modulation goals are possible~(Fig.~\ref{fig:Approach}B). For instance, one intuitively simple goal is to increase the activity of a particular IT target neural site without affecting the other IT sites (henceforth ``cardinal direction modulation''). For this case, each of those non-target sites should ideally respond to every perturbed image in exactly the same way as it responds to its corresponding base image.  
In the more general case, the target modulation direction can be any unit vector in the recorded-IT space with non-negative coefficients. Preliminary work suggested that negative modulation, namely, decreasing neural activity, typically resulted in weaker modulation and was not the focus of this study~(fig.~\ref{fig:SignedCardinal}). Because the total number of possible target modulation directions is infinite, we here tested~(\mytexttilde60) arbitrarily chosen ones, with the goal of determining the central tendency and distribution of what can currently be achieved with this technology.  

\subsection*{Standardizing measurements of directional neural population modulation}

Because neural sites are diverse in response range and exhibit intercorrelated responses, we define and operate in the canonical, identity-covariance IT space. This neutralizes those two idiosyncrasies and allows us to measure neural activity modulation in general operating range~($\sigma$)~units, enabling quantitative comparison and averaging over arbitrary IT subpopulations~\cite{methods}.
Furthermore, to allow for comparability of the intended modulation effects~(DM) with the side effects, we scale the measure of non-intended side effects~(DOM) such that it is neutral with respect to the space dimensionality, rendering its magnitude directly comparable with the DM, regardless of population size. In tasking the model to achieve each modulation goal, we used DOM budgets of $0.1-0.5\sigma$~[see~\cite{methods} for details].

\subsection*{What do current models of the ventral stream predict is possible?}
Here we focus on one particular contemporary ANN model of the ventral stream that has been shown to be quite accurate~\cite{methods}.  We started by asking the model whether there might exist, for a given intended modulation direction, a \emph{fixed} perturbation that could be added to \emph{any} base image to achieve that modulation. Those computational experiments 
show that this generally results in zero net predicted Direction Modulation effects~(DM; supplementary text, fig.~\ref{fig:RoseGlasses}). This prediction is likely a consequence of the highly non-linear causal relationship between the pixel space, where the perturbation is added, and the IT population space -- a non-linear relationship that an accurate model should at least implicitly capture. 

In contrast, allowing the model to search for perturbations that are base-image contingent yielded perturbations that are typically predicted to achieve substantial effects in the intended modulation direction, of order $\text{DM}=2.5\sigma-8\sigma$, while maintaining a small DOM magnitude~(fig.~\ref{fig:MainResults_Detailed}).
Intuitively, the model is telling us that, for \emph{any} base image, if given a small, but non-zero DOM budget~\cite{methods}, it can almost always find image perturbations that it expects will modulate neurons at magnitudes~(DM) as large or larger than each neuron's typical min-to-max response difference~[see operating range in~\cite{methods}] with minimal effects on non-targeted neurons.
For comparison, we also tested common off-the-shelf image enhancement methods, which are typically image-dependent, and we found that these were predicted to be largely ineffective for precise neural modulation~(supplementary text, fig.~\ref{fig:PhotoshopEnhance}).

Interestingly, the model also predicted that the strength of predicted neural modulation~(i.e., the predicted DM) will depend quite strongly on the intended modulation direction, with DM predictions spanning approximately $4\sigma-10\sigma$ over the sample of intended directions~(here, averaging the predicted DM over all sampled base images within each direction).
Motivated by these predictions, we focused our animal experiments on testing the neural effects of model-guided, base-image-contingent perturbations.

\begin{figure}[t]
  \centering
  \includegraphics[scale=1.2]{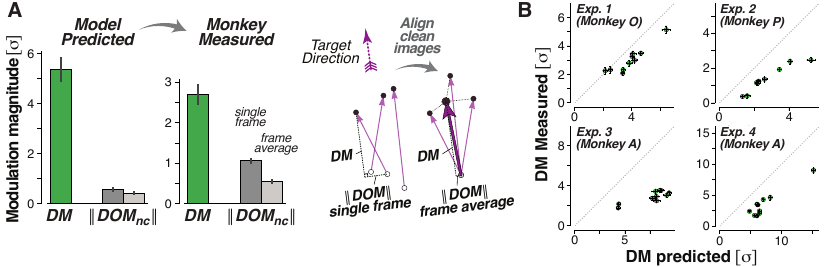}
  \vspace*{-.3cm}
  \caption{
  \textbf{Predicted and measured directional IT neural modulation along experimenter-chosen general modulation directions.}
  {\small {\it
  (\textbf{A})~The modulation along the target direction~(DM, higher is better) is several-fold larger than the direction orthogonal modulation magnitude~(DOM, lower is better). Data are averaged across four control experiments from three different animals, each involving ten pre-selected non-cardinal directions in a ten-neuron space, 50 base images, and \mytexttilde40 presentation repeats. We report noise-corrected DOM scores for both single-frame frame-average cases~(see geometric interpretation)
  . Error bars, SEM over target directions by bootstrap. DOM scores are noise-corrected using bootstrap-estimated SEMs over the base and perturbed images~[see~\cite{methods}].
  (\textbf{B})~Each point shows predicted and measured availability of a sampled target modulation direction, averaging results over all tested base images. Panels correspond to the four experiments summarized in~(A). Notably, the model predicts that some target modulation directions are much more ``available'' than others, namely, have a larger predicted DM for a given DOM budget~(fixed in each panel). The rank order of those predictions is largely borne out in the monkey results~(note correlation of each panel).  But the model generally over-predicts the measured DM~(also see panel A).
  }}
  }
  \vspace*{-.2cm}
  \label{fig:MainResults}
\end{figure}

\begin{figure}[t]
  \centering
  \includegraphics[scale=1.2]{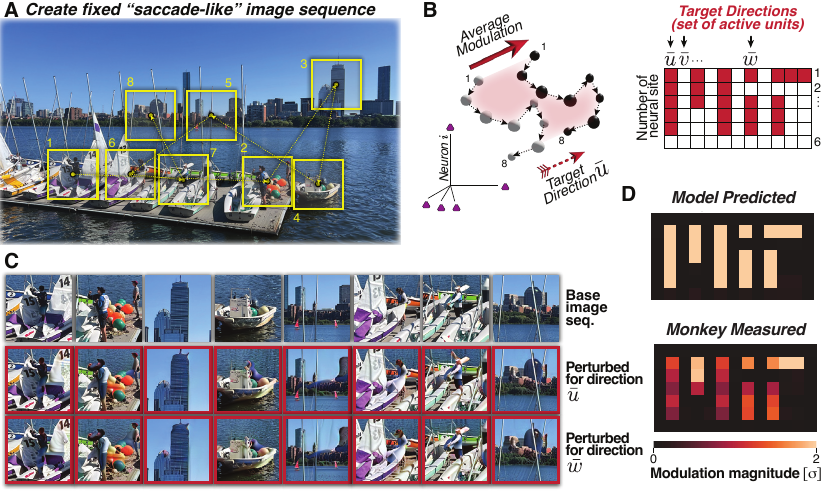}
  \caption{
  \textbf{
  Tests of neural modulation during pseudo, subject-driven, natural visual feeds.
  }
  {\small {\it
  (\textbf{A})~For each of twelve natural base scenes~(one shown here), we created a base sequence of eight cropped images, designed to approximate possible saccadic sampling during natural viewing.  Unlike the ``lab mode'', here the crop images were: (i)~presented nearly back to back for a typical natural viewing fixation duration~(\mytexttilde180~ms), and (ii)~their presentation order was kept fixed within each sequence. We tested eleven modulation directions~(see below), each with twelve base sequences.
  (\textbf{B})~We selected the MIT logo as an arbitrary target modulation direction, decomposed column-wise to eleven different modulation directions within a six-dimensional IT space. 
  (\textbf{C})~Example natural base scene sequence and its perturbed images, designed for two target modulation directions~(B).
  (\textbf{D})~Model predicted and monkey measured average response change~(relative to base sequence responses). Top panel: model predicts near-perfect feasibility of successful injection of the target modulation pattern~(MIT). Bottom panel: recognizable target pattern, as measured in monkey IT neural recordings. Showing average modulation over~96 change vectors~(12 base scenes $\times$ 8 image-crops) measured for each modulation direction, each derived from 25 presentation repeats of each image.
  }}
  }
  \vspace*{-.2cm}
  \label{fig:Demo}
\end{figure}

\subsection*{Empirical testing in monkey IT sub-populations: cardinal direction modulation}
Fig.~\ref{fig:ConceptualResults} shows example results for three directional neural modulation experiments, each corresponding to a different modulation goal.  Each of these goals is a simple cardinal-axis modulation goal in an arbitrarily chosen set of seven simultaneously recorded IT neural sites. In this simple setting, the directional modulation goal was to increase the activity of just one target site~(out of seven), denoted as site 1, site 2 and site 3, respectively, without affecting the natural activity level of the other six IT sites.  Here we tested this for a fixed set of 50 base images.  Specifically, we asked the model to find a different image perturbation for each of the three directional modulation goals~(Fig.~\ref{fig:ConceptualResults}A).  We plot the (Z-scored) responses of each targeted neural site, rank-ordered by its natural response to each base image, and we provide visualization for four of those base images and their perturbed versions. 

Notably, visual inspection suggests that the perturbations found by the model are (i)~structurally diverse, (ii)~strongly dependent on both the start image and the target direction, and (iii)~nearly imperceptible to the naked eye.
Focusing on the target sites alone shows that the perturbed images typically induce strong activity increases -- with average modulation magnitudes of $5.4\sigma$, $3.8\sigma$, and $3.5\sigma$ -- exceeding the site's natural operating range
~(Fig.~\ref{fig:ConceptualResults}B top).
In native units, the measured base-image spike-rate responses are within the ranges, 54-294, 78-223 and 27-164 spikes/s, respectively. In contrast, the perturbed-image responses reach \textit{average} values of 514, 252, and 152 spikes/s~(we note, this is multi-unit activity, see~\cite{methods}). 
The area between the base and perturbed curves reflects the average Direction Modulation~(DM) measured.
To demonstrate precision, we plot the activity \textit{change} relative to the base image responses. This plot reflects the activity modulation for the target site, along with the effects on the other sites in the population, including those targeted in adjacent panels~(Fig.~\ref{fig:ConceptualResults}B bottom). For each of the highlighted sites 1-3, the activity is strongly increased only when the site is specifically targeted, with weak changes in its response when other, nearby sites are targeted~(see color curves). 
Importantly, the response changes~(relative to the normal response to base images) of the non-target sites are typically close to zero. 
In this cardinal modulation case, the changes in all the non-target sites reflect the DOM population vector, which represents the orthogonal hyperplane projections of the modulation.

To connect the specific case of cardinal modulation, which refers to boosting the activity of a single target site, to the more general concept of neural population modulation in any direction, Fig.~\ref{fig:ConceptualResults}C shows plane projections of the results from Fig.~\ref{fig:ConceptualResults}B. Specifically, we visualize projections onto the plane defined by the three cardinal directions considered~(Fig.~\ref{fig:ConceptualResults}C, left panel), and onto the orthogonal planes for each direction~(right panels). 
This visualization shows how the base image distribution~(empty circles) can be consistently and significantly ``shifted'' in the target direction, with average modulation vectors nearly co-linear with the intended direction. When viewed from the direction indicated by the target, we observe some unintended modulation. However, these displacements are significantly smaller in magnitude and not consistent in direction.

\subsection*{Empirical testing in monkey IT sub-populations: general direction modulation}
We tested directional neural modulation in arbitrary, non-cardinal, target directions in three animals. Fig.~\ref{fig:MainResults} provides an overview of four distinct experiments, reporting on the model-predicted and monkey-measured statistics of DM and DOM magnitude. For DOM, we report two variations~(illustrated): the single-frame DOM, as previously defined, and a frame-average version, which computed the DOM based on the average modulation vector. The latter approach allows for balancing out orthogonal projections that vary across images, simulating potential downstream effects of temporal integration. 
These findings suggest that precise neural modulation in IT is generally achievable, with typically observed DM of $2.7\sigma$, exceeding the single-frame and frame-average DOM magnitude scores by 153\% and 393\%, respectively. Notably, the achieved DM and the precision~(i.e., DOM magnitude) are overestimated by the model, suggesting a residual misalignment between the simulated and the biological IT neural populations.
We further used the model to explore target direction ``availability'' -- which refers to the level of modulation achievable for a particular direction under a DOM budget constraint -- and found that it greatly varies depending on the direction~(Fig.~\ref{fig:MainResults}B). That is, some directions have a larger predicted DM for a given DOM budget~(note the spread of dots along the x axis in each panel; DOM budget is fixed in each panel). 
By comparing the monkey-measured and the model-predicted DM for the sampled directions used in the experiments, we found that the rank order of direction ``availability'' is generally predicted by the model, but that the model tends to overpredict the magnitude of the DM that will be achieved. The model's predictive power accuracy varied across experiments, being largely influenced by the specific animal and the reliability of the corresponding neural sites~(see fig.~\ref{fig:ModelAccuracy} for model fit accuracy).

\subsection*{Empirical assessment of applicability during typical natural viewing feeds}
The results above were obtained in standard ``lab mode'': using IID-sampled natural images sourced from a well-known Natural Image dataset, fully-randomized image-by-image, and presented with a 50\% ON/OFF duty cycle as typically practiced in the standard primate RSVP paradigm~\cite{Bashivan2019NeuralSynthesis, Guo2022AdversariallyRepresentations, Dapello2023AligningRobustness,Schrimpf2020IntegrativeIntelligenceb, Ponce2019EvolvingPreferences}. 
We here asked whether the proposed modulation method is likely to generalize to fully natural viewing conditions, wherein the ``natural images'' impinging on the retinae result from saccadic sampling of a larger natural scene~\cite{DiCarlo2000FormViewing}. 
To emulate saccadic exploration of natural scenes, we crafted 1.6-sec image sequences, each consisting of eight selected image crops from each natural scene, as illustrated in Fig.~\ref{fig:Demo}A. Further adhering to natural viewing scenarios, we used twelve randomly chosen natural scenes, which we casually photographed with a smartphone camera. For testing IT neural populations, each image crop was displayed for approximately 180~ms (mimicking a fixation), with 20~ms between image crops (mimicking a saccade), always in a fixed predetermined sequence for each natural scene, intended to give us a reliable estimate of the mean firing rates that would be evoked by a typical, organism-driven saccadic sequence. To assess our modulation approach, we compared IT responses evoked by sequences of completely unmodified image crops~(above) with responses evoked by corresponding sequences of images where all image crops in the sequence were perturbed for a particular modulation goal, described next. 

To assess the possibility of arbitrary, predetermined population modulation, we selected modulation target directions corresponding to the MIT symbol~(Fig.~\ref{fig:Demo}B). Although achieving this pattern at displayed resolution would require 91 IT sites, we opted for a column-wise decomposition due to the limited number of visually reliable sites, thereby defining the pattern via a set of 11 directional modulation goals in a 6-dimensional population space.
Thus, for each base scene, we created and presented, in addition to the base sequence, eleven fixed sequences of eight images, corresponding to each IT modulation goal.  All sequences were randomly interleaved during the experiment.

As in previous experiments, we used the image-computable model to generate perturbed image sequences for each target direction~(Fig.~\ref{fig:Demo}C). Visualized as a heatmap of population response change, these perturbations were predicted to produce a very good rendition of the MIT pattern in the image-mean modulation. Consistent with this prediction, IT neural population recordings revealed a clearly recognizable MIT modulation pattern.
We found no significant qualitative or quantitative differences between the pseudo-natural viewing mode and the ``lab'' mode, with typical DM values computed in a comparative analysis, of $2.9\sigma$ and $3.2\sigma$, respectively.  
These results suggest that strong neural population modulation can, in principle, be accomplished during natural viewing, provided that the expected incoming image is provided at a latency short enough to derive each perturbation.

\subsection*{Unlocking vision-based noninvasive neural interventions using sensory-computable models of visual processing}
Our results show that current machine-executable models of the ventral stream are now accurate enough to induce vision-based activity changes in neural populations -- congruent with predetermined, experimenter-chosen target patterns -- in the highest level of ventral visual representation, the IT cortex. While the precision achieved is not perfect, we found strong directional neural modulation along the intended direction, which drives IT to rarely observed population activity states~\cite{Batista2019}, can be achieved using model-guided image perturbations applied to examples of natural visual feed. We report a measured neural modulation efficiency, i.e., DM to DOM-magnitude~(DOMM) ratio, of~2.6:1 and~4.9:1 for the single-frame and frame-average cases respectively, averaged across animals, experiments, start images, and arbitrary target directions.

What limits the current modulation efficiency? 
First, 
the existence of response correlation between some neural sites suggests that a given target site is not free to be modulated without at least partially affecting those other sites ~[see ``intrinsic manifold'' in~\cite{Sadtler2014,Jazayeri2017}].  Indeed, we observe off-target effects despite our attempts to neutralize this by working in the canonical space. This is also evident in the computational model of the neurons: We found that imposing a zero DOM budget constraint made it impossible to identify modulation-inducing perturbations. A finite DOM budget was thus required by the model, predictively setting an upper bound of less than 10:1 on modulation efficiency of mapped neural sites.

Second, we observed that the model overpredicts the modulation efficiency relative to the observed neural efficiency~(Figs.~\ref{fig:MainResults},\ref{fig:Demo}).  
This argues that (i)~each animal has idiosyncrasies in its ventral stream that cannot be captured by any single backbone model, or that (ii)~even if all biological ventral streams are identical, the model backbone we used is not a veridical hypothesis of that ventral stream, or that (iii)~the backbone model neurons were not accurately mapped to the biological neural sites, or some combination of all three. 
While disentangling these ideas is beyond the scope of this study, we did computer simulations to explore the effect of the mapping procedure (iii) on modulation efficiency in a digital monkey surrogate, generated as sampled units from our backbone model. We found that, even with mapping data at the scale of ImageNet, our simulations show a five-fold drop in observed modulation efficiency, relative to predicted efficiency. 
Interestingly, analogous challenges have been explored in artificial neural networks and the limited transferability of their predictions~\cite{Geirhos2020ShortcutNetworks}. Future studies are needed to quantify the relative impacts of these current modeling shortcomings (i, ii, iii) on artificial and biological modulation efficiency, and to explore its theoretical upper bound.

Notably, we found that the direction of the~(unintended) orthogonal projection of the induced neural population modulation~(i.e., the direction of the DOM vector, Fig.~\ref{fig:ConceptualResults}C) varies with the start image.  This is  
indicated by the mean DOM magnitude being lower in the frame-average case (where the DOM vectors pointed in different directions can cancel out) as compared with the single-frame case~(Fig.~\ref{fig:MainResults}A). This suggests that, if brain regions downstream of IT have timescales of temporal integration that are longer than a typical natural fixation duration~(\mytexttilde200~ms), then the effective downstream neural modulation efficiency in those regions is expected to be higher than reported here. 
In our experiments, this efficiency nearly doubled when averaging across 5-10 images, akin to an integration timescale of  1-2~s.

Harnessing neural population control for potential internal cognitive state modulation may require consideration of the entire IT population -- which is approximately 10 million neurons with an unknown effective intrinsic dimensionality~\cite{DiCarlo2012HowRecognition}, but likely at least 1000~\cite{Lehky2014} or more~\cite{Gauthaman2024TheCortex,Gauthaman2024UniversalCortex}. The strength and efficiency of neural modulation reported here most safely pertain to the smaller population-size regime considered. Nevertheless, this is the first time that population-level precise neural modulation has been measured in any size population in IT. An important future direction is to test the scaling of this approach and to test its predicted and empirical impact on perceptual state measures~(i.e., implicit and explicit behavioral report measures).

While not the goal or focus of this study, an interesting property of the model-identified perturbations is that they appear to be nearly imperceptible, despite being neurally very potent~(Fig.~\ref{fig:ConceptualResults},\ref{fig:Demo}). This suggests the notion of neural code \textit{degeneracy}, where many IT states may correspond to a similar perceptual state. In principle, machine-executable models can be used to explore degenerate spaces: degrees of freedom in high-level neural representation of stable percepts~[cf.~Metamers~\cite{Freeman2011}].
This observation is consistent with previous reports on the sensitivity of the IT representation and its supported behaviors to small norm natural image perturbations~\cite{Guo2022AdversariallyRepresentations, Gaziv2023StrongANNs,Talbot2024L-WISE:Enhancement,Elsayed2019AdversarialPerception}. 
Intriguingly, from an application point of view, depending on how a degenerate IT representation relates to directly connected downstream areas, such as the prefrontal cortex or the amygdala~\cite{Kravitz2013, Stefanacci2002}, it may be that precise, strong neural modulation in those target brain regions could be induced by perturbations designed with the methods used here, while also possibly having minimal interference with normal, ongoing visual perception and behavior. 
While prospective and speculative, this opens an avenue for beneficial applications of brain states modulation for humans, harnessed through noninvasive -- and \textit{non-interrupting} -- subtle perturbations applied to natural visual feed. Such perturbations could potentially be delivered via low-latency augmented reality glasses, or by augmenting any otherwise screen-delivered visual stimulus.

\subsection*{Materials and Methods}

\subsubsection*{Electrophysiological recordings in macaques}
We sampled and recorded neural sites across the Rhesus macaque~(\textit{Macaca mulatta}) IT cortex, including its posterior, anterior, and central sub-regions in three awake, behaving macaques. In each monkey, we implanted 2-3 chronic Utah arrays with 96 microelectrodes each (1.5~mm, 400~µm pitch, Iridium Oxide-coated electrodes; [see~\cite{Neurotech}]). The neural sites considered in this study generally record multiple local neuronal responses per channel, referred to a multi-unit activity.

\subsubsection*{Neural site inclusion criteria \& visual reliability}
We selected neural site populations for experiments mainly based on their visual reliability. We computed the single-repeat reliability (SRR). The SRR is the single-repeat Pearson correlation over randomly sampled repeat pairs, corrected for small samples. Sites with SRR $\geq 0.3$ were deemed valid for analysis, which is consistent with criteria used in previous studies~\cite{Bashivan2019NeuralSynthesis}. For some animals, a threshold of 0.4 was used. This gave rise to populations of approximately 10 sites per experiment. In analyzing cardinal modulation, we also employed minimal guidance on subset selection to ensure reduced pairwise correlation.

\subsubsection*{RSVP experimental protocol}
For testing neural responses to each image, we used a standard Rapid Serial Visual Presentation~(RSVP) paradigm, with image durations approximating those of natural viewing, using MWorks~\cite{MWorks}.
All images were presented at $8^{\circ}$ over the visual field. Monkeys fixated a white circular dot (0.2°-0.3°) for 300~ms to initiate a trial. On each behavioral trial, we presented a sequence of eight randomly sampled images, each ON for 100~ms followed by a 100-ms OFF, showing a gray blank screen. Successful fixation for the entire trial was followed by a water reward and an intertrial interval of 500~ms, followed by the next sequence. Trials were aborted if gaze was not held within $\pm$2.5°-4° of the central fixation dot during any point. 
To eliminate any adaptation effects, all tested images~(base and perturbed) were fully randomly interleaved, and each was typically presented 40 times.
For the pseudo natural viewing condition experiments~(Fig.~\ref{fig:Demo}), we aimed to simulate the saccadic duration between each sampled location in the scene. We thus adjusted the ON/OFF durations to 180/20~ms and imposed a fixed predetermined order of the images \textit{within} sequences while the~(fixed-order) sequences were randomly interleaved, one per behavioral trial~(above). 

\subsubsection*{Neural data preprocessing}
For spike detection, we applied 3$\sigma$ thresholding over the voltage waveform of each multi-unit channel.
Behavioral~(eye-tracking \& MWorks events) and neural data were time-registered using a photodiode signal co-presented with the stimuli at the screen's corner.
The stimulus-paired neural response was defined as the mean spike rate in the 70-170~ms window from stimulus onset, consistent with previous IT studies~\cite{Hung2005FastCortex,Kar2021FastRecognition,Guo2022AdversariallyRepresentations,Dapello2023AligningRobustness,Majaj2015SimplePerformance}.

\subsubsection*{Natural Image datasets}
We used the ImageNet~\cite{Deng2009ImageNet:Database} validation set as the main Natural Image dataset for model mapping and for the main neural modulation experiments, excluding the natural viewing experiments. We refer to these images as ``base images''~(in the adversarial literature, the term ``clean'' is typically used).
Standard ImageNet transforms were used to convert the images to~\mbox{$224\times224\times3$} RGB pixel space~(\mbox{$D=150,528$}). 
For the application demo, we used as ``base scenes'' a randomly selected set of large images captured by a Google Pixel~5 smartphone. To define image sequences, we manually sampled non-overlapping image crops that mimic a possible natural scene exploration via a series of typical length saccades with intervening fixations.

\subsubsection*{Modeling an IT population}
To model an IT neural population, we used an ImageNet-pretrained ``robustified'' (i.e., Adversarially-Trained) ResNet-50 feature extractor, which is among the currently leading models of the ventral visual stream~\cite{Guo2022AdversariallyRepresentations, Gaziv2023StrongANNs}. Consistent with ~\cite{Guo2022AdversariallyRepresentations}, which focused on IT representation sensitivity, we used the $\varepsilon_{tr} = 2.0$ model variant and designed a linear mapping from the high-level representation in layer 4.0 to the number of units in the population of interest. This mapping was randomly initialized.

To fit the mapping layer to the specific population, we collected neural responses to natural images, forming a mapping set. The image set consisted of 10,000-11,000 distinct base images sampled uniformly from ILSVRC~\cite{ILSVRC} object categories, and an additional 1,000 images used for model validation. Two repeats were recorded for each image (totaling 20,000-22,000 image presentations for the session, excluding normalizer sets).

We optimized the model parameters on the image and mean-centered single-repeat response pairs as the training data for 10 epochs using Adam, an MSE objective, a learning rate of $10^{-3}$ with 0.6 exponential decay, with early stopping based on the validation set. While in principle trained end-to-end, we adjusted the learning rate applied to the pretrained backbone to be a tenth of the global learning rate; we found the results to be insensitive to this minimal finetuning of the backbone. We applied $\ell_{2}$ and $\ell_{1}$ regularization on the mapping layer's weights to encourage weight sparsity and magnitude compression.

The resultant model defined the image-computable simulated IT population.  This was then fixed and used for guiding the design of the image perturbations for a given target modulation. Fig.~\ref{fig:ModelAccuracy} reports on the obtained model accuracy (raw and noise-corrected) for each experiment.

\subsubsection*{Generating perturbed images to modulate neural activity} 
Following \cite{Gaziv2023RobustifiedPercepts, Guo2022AdversariallyRepresentations, Madry2018TowardsAttacks, Talbot2024L-WISE:Enhancement}, we used Projected Gradient Descent~(PGD) to perturb images for a range of predetermined neural modulation goals imposed on the model output. Importantly, we defined the modulation goals on the canonical output space [i.e., identity-covariance, see~\cite{methods}]. The mapping model between the image and the canonical simulated IT space with $n_{s}$ neural sites reads, \mbox{$f:\;\mathbb{R}^{H\times W\times3}\rightarrow\mathbb{R}^{n_{s}}$}. Given a start image $x$, we optimize for the image perturbation, $\delta\in\mathbb{R}^{H\times W\times3}$, that minimizes the objective on the output change, namely,

\vspace*{-.4cm}
\begin{equation}\label{eq:Objective1}
    \delta^{*}=\operatorname*{argmin}_{\delta}{\cal L}\left(\vec{\Delta}\left(\delta\right);\;\hat{u},b\right)\;\;;\vec{\Delta}\left(\delta\right)\equiv\vec{f}\left(x+\delta\right)-\vec{f}\left(x\right),
\end{equation}
\vspace*{-.3cm}

\noindent
where $\hat{u}$ and $b$ are the target direction and the DOM budget. The objective consists of a DM maximization term and a DOM magnitude budget-controlled term,

\begin{equation}\label{eq:Objective2}
    {\cal L}\left(\vec{\Delta}\left(\delta\right);\;\hat{u},b\right)=\underbrace{\left(1-\frac{\vec{\Delta}\cdot\hat{u}}{t_{DM}}\right)^{2}}_{{\cal L}_{DM}}+\lambda_{\text{orth}}\underbrace{R\left(\left\lVert P_{\hat{u}}^{\perp}\vec{\Delta}\right\rVert _{p}-b\right)}_{{\cal L}_{DOM}},    
\end{equation}

\begin{equation}\label{eq:DOM}
    P_{\hat{u}}^{\perp}\vec{\Delta}\equiv\frac{\vec{\Delta}-\left(\vec{\Delta}\cdot\hat{u}\right)\hat{u}}{\sqrt{n_{s}-1}},
\end{equation}

\noindent
where $t_{DM}=10^{3}$ is an arbitrarily large DM target, \mbox{$P_{\hat{u}}^{\perp}\vec{\Delta}$} is the DOM, i.e., the modulation projection to the hyperplane orthogonal to direction $\hat{u}$, normalized by the hyperplane dimension (for magnitude comparability with the DM); $R\left(\cdot\right)$ is the ramp function, gating the penalty on DOM magnitude by the given budget. We used $p=2$ in our experiments, corresponding to $\ell_{2}$-norm of the DOM vector. $\lambda_{\text{orth}}=10^{6}$ is an arbitrarily large penalty coefficient.

The optimization is performed in gradient descent steps: 

\vspace*{-.3cm}
\begin{equation}\label{eq:PGD}
    \delta_{k+1}\leftarrow\delta_{k}+\eta\nabla_{\delta}{\cal L}\left(\vec{\Delta}\left(\delta\right);\;\hat{u},b\right),
\end{equation}
\vspace*{-.4cm}

\noindent
where $k$ denotes the step, and $\eta$ is the step-size. We typically performed 5,000 steps with a step-size of 0.01.

\subsubsection*{Selecting non-cardinal directions}
The number of directions generally scales like the area of $n_{s}$-dimensional sphere, $n_{s}$ being the population size. To sample from this space, we focused on composite-cardinal directions of degree $k$: directions with `1' in up to $k$ elements, with zeros for the rest, normalized to unit length. Given typical populations of size 10, we used $k=5$, spanning a pool of 637 directions that were all evaluated for their ``availability'' (see Main Text) using the model and a held-out set of 50 natural images.
For experiments 1-3 (Fig.~\ref{fig:MainResults}B) we selected the eight most and the two least predictively available directions. And for experiment 4 we uniformly sampled 10 directions from the availability-ranked pool of directions.

\subsubsection*{Selecting DOM budgets}
Preliminary analysis revealed that models of any given neural population predict an intuitive trade-off between the achievable DM and the DOM budget: Allowing a model to have larger predicted DOM magnitude~budgets~(predicted side-effects) also allowed it to, in general, find perturbations that also led to larger predicted DM scores~(desired effects; fig.~\ref{fig:DM_DOM_Tradeoff}).  
For this study, we thus focused on cardinal and general directional modulation, using a specific regime of DOM budgets, in the range $0.1-0.5\sigma$. 
For intuition, this means that, for cardinal modulation experiments, we were asking the model to restrict the predicted modulation of non-targeted sites to no more than 5-25\% of each site's natural operating range~(2~SDs).

\subsubsection*{Noise corrections}
Non-reproducible variance in the measured neural responses will inflate the measured DOM magnitude, even if the modulation is perfect.  As such, we carefully corrected for that, with tests of these procedures to insure that we were not over-correcting.  Specifically, we computed the SEM over the repeats by bootstrap in both the base and the perturbed image responses. We then subtracted the orthogonal plane projections of their squares from the squared raw DOM (variance units). For the frame-average case, we followed a similar procedure, but also averaged across stimuli in each bootstrap sample.

\subsubsection*{Canonical space transforms}
The scaling transformation enabled us to have a meaningful interpretation and comparability of the neural activity modulation results that are pooled across a range of specific neural sub-populations and modulation directions. %
In this transformed neural space, one sigma unit is the standard deviation of the normal visually-driven operating range of each neutral site, measured over a set of approximately 50 natural images~(to scale recorded responses). Intuitively, $4\sigma$ spans the 95\% operating range, or approximately the min-to-max response difference.

In detail, we defined the main canonical space to be the zero-mean, identity-covariance IT space for the simulated and measured responses, respectively. We followed a standard linear whitening/sphering protocol based on eigenvalue decomposition, which was computed per experiment on a bag of base-image responses. For the measured IT space, we used the repeat-averaged neural responses for the base-image subset, and for the model's output transformation, we used predicted responses for a bag of 10,000 base natural images from ImageNet, uniformly sampled from the 1000 object categories. See supplementary text and fig.~\ref{fig:CanonicalTransform} for geometric interpretation. 

Notably, some of the modulation experiments were pursued using a ``cardinal'' normalization transformation, which scales the responses of each site by its own operating range without accounting for inter-site correlations. For uniformity, all the results presented followed the main canonical transformation advocated, despite being suboptimal with respect to the modulation goals practically employed for some experiments~(experiments 1-2, Fig.~\ref{fig:MainResults}B). We found the results and the deduced conclusions to be largely insensitive to this design and analysis choice~(see figs.~\ref{fig:MainResults_CardinalNorm},\ref{fig:MainResults_Detailed}).


\bibliographystyle{sciencemag}

\section*{Acknowledgments}
We thank C.~Guo and M.~Lee for insightful discussions, and C.~Shay for administrative support.

\paragraph*{Funding:}
This work was partially funded by the Office of Naval Research (N00014-20-1-2589, JJD), and the Simons Foundation (NC-GB-CULM-00002986-04, JJD).

\paragraph*{Author contributions:}
G.G. and J.J.D. designed research; G.G. conducted analysis, performed image generation for neural modulation, and analyzed neural data; Y.B., S.G., A.AH., and J.J.D. performed animal surgeries; S.G. and A.AH. performed neural recordings; G.G. and J.J.D. wrote the paper.

\paragraph*{Competing interests:}
J.J.D. is a member of the scientific advisory boards of The Wu-Tsai Institute at Yale University, The Lefler Center at Harvard University, and The ARNI Center at Columbia University.

\paragraph*{Data and materials availability:}
We make our code and data publicly available at \\
\texttt{https://github.com/ggaziv/DirectionalNeuralModulation} upon publication.

\newpage

\renewcommand{\thefigure}{S\arabic{figure}}
\renewcommand{\thetable}{S\arabic{table}}
\renewcommand{\theequation}{S\arabic{equation}}
\renewcommand{\thepage}{S\arabic{page}}
\setcounter{figure}{0}
\setcounter{table}{0}
\setcounter{equation}{0}
\setcounter{page}{1} %

\begin{center}
\section*{Supplementary Materials}
\end{center}

\subsection*{Supplementary Text}

\subsubsection*{Model predictions for image-fixed perturbations}
The notion of image-agnostic perturbations is reminiscent of the notion of ``rose-colored glasses'', albeit it has more degrees of freedom in the pixel changes it can support. From a scientific but also an application point of view, we sought to analyze whether image-specific updates are necessary -- or excessive -- for supporting strong and precise neural modulation.
To analyze the possibility of inducing directional modulation using image-agnostic perturbations, we incorporated a regularization objective that enables fine and graded control over the uniformity level of the resulting image perturbations for a bag of base images. Given a batch of images of size $K$, we optimize for perturbations $\left\{ \delta^{(k)}\right\} _{k=1}^{K}$, and penalize perturbation divergence, which encourages a fixed perturbation:

\vspace*{-.2cm}
\begin{equation}\label{eq:RoseGlasses}
{\cal L}_{FP}=\frac{\mathbb{\mathbb{E}}_{k}\left[\left\lVert \delta^{(k)}-\bar{\delta}\right\rVert _{2}\right]}{\left\lVert \bar{\delta}\right\rVert _{2}}.
\end{equation}
\vspace*{-.3cm}

\noindent
We hyperparameterized this loss by a control coefficient, $\lambda_{FP}$, and repeated the model-based analyses and image generation using this auxiliary criterion and a mapped-model from experiment~2.
Fig.~\ref{fig:RoseGlasses} shows the results for varied levels of perturbation fixation levels. Notably, increasing the value of $\lambda_{FP}$ drastically decreases -- up to complete abolishment -- of the predicted DM. To ensure proper calibration for the $\lambda_{FP}$ values used, we computed additional metrics, including perturbation divergence and DM scores predicted for the average perturbation; indeed, the perturbations obtained under this range of values were not over-regularized.
This experiment suggests that image-contingent perturbations are necessary to unlock precise neural modulation.

\subsubsection*{Model predictions for off-the-shelf image enhancement methods}
\begin{figure}[t]
  \centering
  \includegraphics{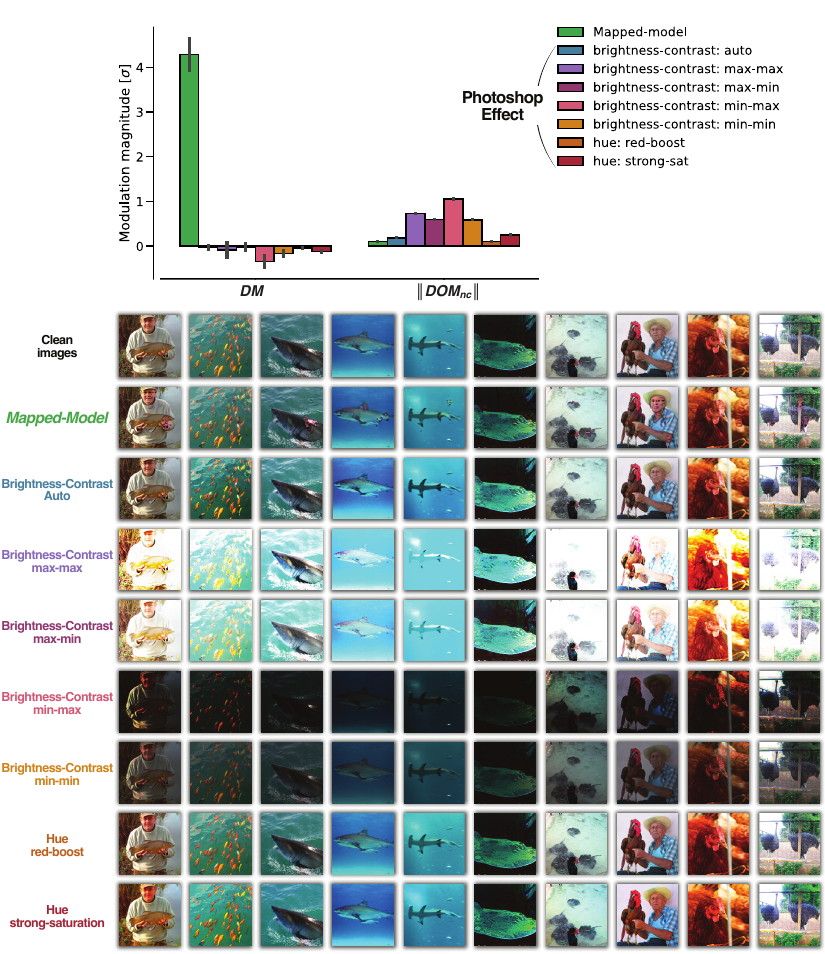}
  \caption{
  \textbf{Off-the-shelf enhancement methods do not generally induce directional neural modulation.} 
  {\small {\it
  We compare the directional neural modulation effects obtainable by model-guided perturbations with those obtainable by widely used off-the-shelf image enhancement methods.
  Showing example enhanced images and mapped-model predictions for seven Adobe~Photoshop-based enhancement effects, and comparing those with the model-based image perturbation effects. Error bars, SEM over the target directions. A set of 50 randomly selected base natural images from ImageNet were used for analysis.
  }}
  }
  \label{fig:PhotoshopEnhance}
\end{figure}

We sought to compare the precise neural modulation effects obtainable by model-guided perturbations with those obtainable by widely used off-the-shelf image enhancement methods.
To this end, we used Adobe~Photoshop~\cite{AdobePhotoshop} software and hand-designed seven image enhancement actions based on standard effects offered. Notably, many of these effects account for local image statistics and are therefore image-dependent~(cf. fixed perturbations, see supplementary text and Fig.~\ref{fig:RoseGlasses}). 
We used the resultant ``enhanced'' images as ``perturbed'' images for subsequent model-based analysis. 
Fig.~\ref{fig:PhotoshopEnhance} shows a comparison of predicted modulation effects by Photoshop-enhanced images with those obtainable by model-perturbed images.
Our analysis suggests that, beyond the Photoshop enhancements being inherently agnostic of any target direction considered, their induced modulation effects are generally much lower in magnitude compared with those induced by the mapped-model, while inducing generally larger DOM effect.

\subsubsection*{Interpreting the canonical space transform}
\begin{figure}[t]
  \centering
  \includegraphics{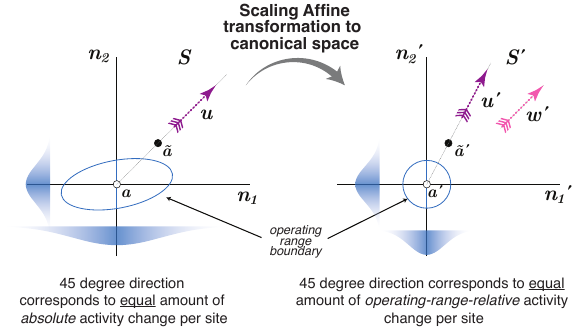}
  \caption{
  \textbf{Geometric interpretation of the canonical space transform.}
  {\small {\it
  $S$ and $S'$ represent the native and the scaled (i.e., transformed) coordinate systems. We consider two selected neurons/sites, with a single pair of base and perturbed image responses, where the base image is centered at the origin.
  }}
  }
  \label{fig:CanonicalTransform}
\end{figure}

Fig.~\ref{fig:CanonicalTransform} provides an illustration of the canonical space transformation used.
In this visualization, $S$ and $S'$ represent the native and the scaled (i.e., transformed) coordinate systems. 
The two neural sites visualized, $n_{1}$, $n_{2}$, differ in their natural operating range~(i.e., SD), while also being slightly correlated.
The canonical space transformation gives rise to new site coordinates, each with a unity natural-operating range. 
In addition, the correlation between the two sites in the native space~(nonzero off-diagonal covariance elements) is eliminated in the transformed space.

Notably, the interpretation of the target directions and directional modulation is strongly dependent on the coordinate space in which they are defined. For example, an activity modulation of $45^{\circ}$ angle in the native space means an equal amount of absolute activity change per site, whereas the transformed version of this direction, $u'$, would generally be of an angle different from $45^{\circ}$. In contrast, in the canonical space, the $45^{\circ}$ direction ($w'$) would correspond to an equal amount of operating-range fold-change per site.

\begin{figure}[h]
  \centering
  \includegraphics{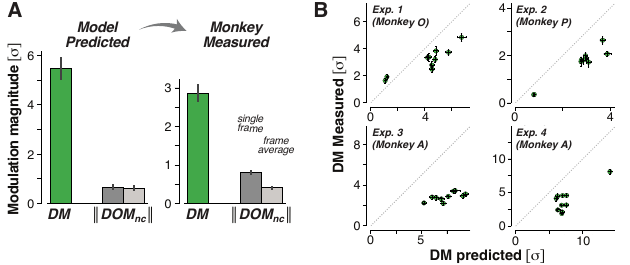}
  \caption{
  \textbf{[Different normalization] The leading models of the ventral stream can support directional neural modulation along experimenter-chosen directions in IT space.} 
  {\small {\it
  Similar to results from Fig.~\ref{fig:MainResults} but re-analyzed using ``cardinal'' normalization, instead of the preferred identity-covariance canonical space normalization. Notably, the cardinal transformation was also used to design the perturbed images in experiments 1-2 (but not for experiments 3-4, see Fig.~\ref{fig:MainResults_Detailed}).
  }}
  }
  \label{fig:MainResults_CardinalNorm}
\end{figure}

\begin{figure}[h]
  \centering
  \includegraphics{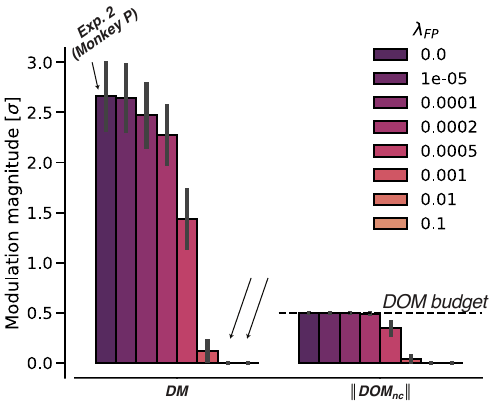}
  \caption{
  \textbf{Models predict that image-fixed perturbations cannot induce strong and precise neural modulation.} 
  {\small {\it
    We repeated the model-based analysis for experiment~2~(Fig.~\ref{fig:MainResults_Detailed}) using an auxiliary perturbation fixation criterion. Error bars, SEM over target directions. A set of 50 randomly selected base natural images from ImageNet were used for analysis. The DOM budget used in this experiment is set to 0.5. 
    }}
  }
  \label{fig:RoseGlasses}
\end{figure}

\begin{figure}[t]
  \centering
  \vspace*{-.5cm}
  \includegraphics{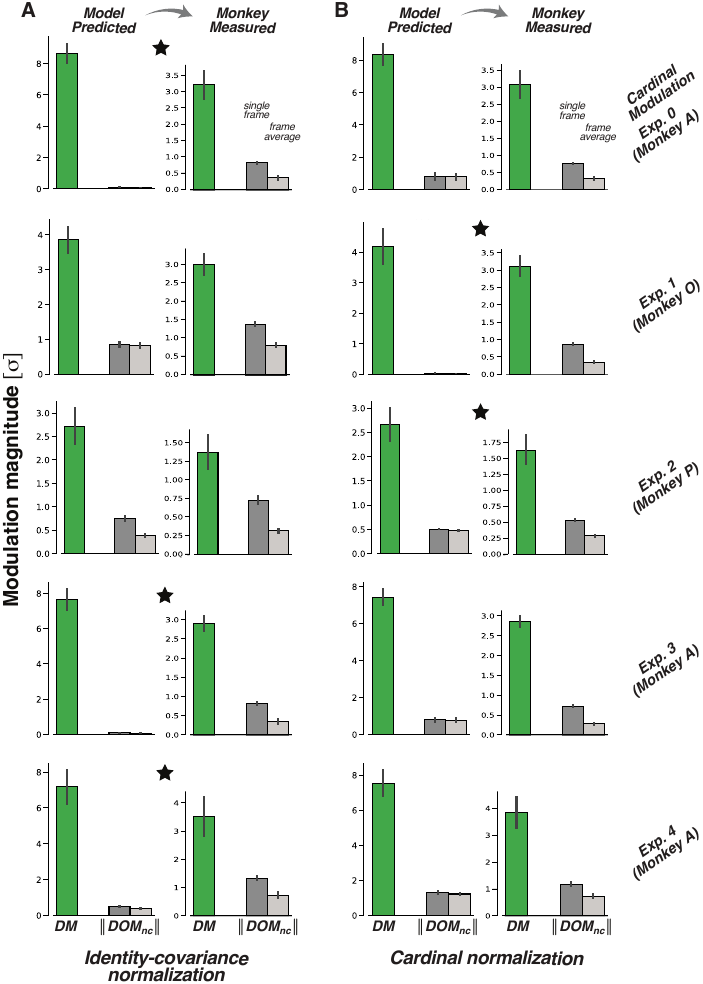}
  \vspace*{-.3cm}
  \caption{
  \textbf{[Detailed view] The leading models of the ventral stream can support directional neural modulation along experimenter-chosen directions in IT space.} 
  {\small {\it
  Detailed version of the summarized results from Fig.~\ref{fig:MainResults}A and a comparable version of the cardinal modulation results from Fig.~\ref{fig:ConceptualResults}. Each row on the left corresponds to a single experiment from the summarized plot in~Fig.~\ref{fig:MainResults}A, or the cardinal modulation experiment. The right column provides similar analyses but using the cardinal normalization (non-multivariate) when analyzing the data, instead of the preferred identity-covariance canonical space normalization. For every experiment, we denote with a star the analysis type where the normalization used for image perturbation matches that used for the modulation analysis (i.e., the normal case for that experiment). 
  }}
  }
  \label{fig:MainResults_Detailed}
\end{figure}

\begin{figure}[t]
  \centering
  \includegraphics{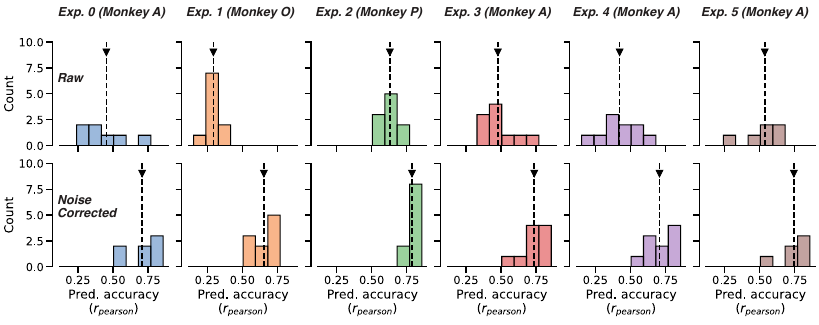}
  \caption{
  \textbf{Mapped-models accuracy.}
  {\small {\it
  We evaluated model accuracy on a held-out set of 1000 base natural images.
  We report accuracy distributions over the predicted output sites for the six mapped-models used in the main text presented experiments: \mbox{Exp. 0}~(cardinal modulation, Fig.~\ref{fig:ConceptualResults}); \mbox{Exp. 1-4}~(four model underlying results in Fig.~\ref{fig:MainResults}); and \mbox{Exp. 5}~(Demo in Fig.~\ref{fig:Demo}). Dashed lines mark distribution averages; the noise-corrected site-average accuracies for these models are: 0.71, 0.65, 0.79, 0.74, 0.71, and 0.74, respectively.
}}
  }
  \label{fig:ModelAccuracy}
\end{figure}

\begin{figure}[t]
  \centering
  \includegraphics{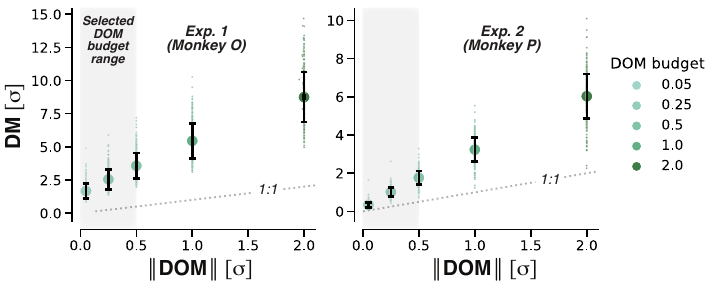}
  \caption{
  \textbf{Models predict a trade-off between neural DM and DOM budget.} 
  {\small {\it
  Increasing tolerance for DOM allows us to increase the typical strength of the intended DM effect. 
  We used mapped-models to evaluate direction ``availability'' -- the typical (i.e., image-mean) predicted DM achievable given a DOM budget -- under varied DOM budget levels. This was done for every direction in a pool of 637 composite-cardinal directions of degree 5 in a 10 dimensional space~(see supplementary text).
  Showing target-direction mean scores and distribution. Error bars, SD over the directions. Shaded area indicates our DOM budget range of choice for monkey experiments. A set of 50 base natural images randomly selected from ImageNet was used as start images.
  }}
  }
  \label{fig:DM_DOM_Tradeoff}
\end{figure}

\begin{figure}[t]
  \centering
  \includegraphics{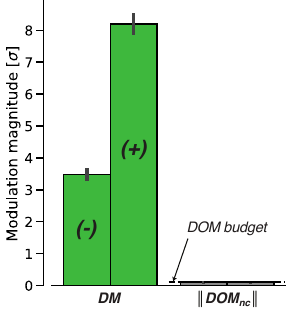}
  \caption{
  \textbf{Comparison of mapped-model predictions for positive and negative cardinal modulation goals.} 
  {\small {\it
  Positive cardinal modulation resulted in larger typical modulation compared with negative (i.e., attempts to decrease the activity), hence was the focus of this study.
  DOM budget of $0.1\sigma$, 10 randomly sampled base start images, 7 cardinal directions (per polarity) in a 7 neural-site population.
  }}
  }
  \label{fig:SignedCardinal}
\end{figure}

\clearpage %

\end{document}